\documentclass[twocolumn,amsmath,amssymb]{revtex4}

\usepackage{natbib}
\usepackage{bibentry}
\usepackage{graphicx}
\usepackage{dcolumn}
\usepackage{bm}
\newcommand{\bra}[1]{\langle#1|}
\newcommand{\ave}[1]{\langle#1\rangle}

\newcommand{\ket}[1]{|#1\rangle}

\begin{document}

\title{Long-time Protection of Nonlocal Entanglement}

\author{Bruno Bellomo}
\author{Rosario Lo Franco}
\email{lofranco@fisica.unipa.it}
 \homepage{http://www.fisica.unipa.it/~lofranco}
\author{Giuseppe Compagno}
\affiliation{CNISM and Dipartimento di Scienze Fisiche ed Astronomiche,
Universit\`{a} di Palermo, via Archirafi 36, 90123 Palermo, Italy}

\date{\today}

\begin{abstract}
We investigate how nonlocal entanglement, as identified by violations of a Bell inequality, may be protected during the evolution. Our system consists of two qubits each embedded in a bosonic reservoir evolving independently and initially in an entangled mixed state. We show that the violation of the Bell inequality can be related to the single-qubit population of excited state in such a way that, by appropriately choosing structured environments that give rise to sufficiently high values of population trapping, long-time protection of nonlocal entanglement can be correspondingly achieved.
\end{abstract}

\maketitle

\section{\label{intro}Introduction}
Entanglement is one of the fundamental but counterintuitive aspects of quantum mechanics\cite{epr}. A basic trait of entangled, or quantum correlated, systems is that a measurement on one subsystem of the composite system determines the state of the other, independently on their separation. Entangled systems can also be characterized by the presence of quantum correlations that cannot be simulated by any classical local model\cite{bell}. However, while for pure states the presence of entanglement always corresponds to the presence of nonlocal quantum  correlations (NLQCs)\cite{gisin1991PLA,gisin1992PLA}, this is not the case for mixed states; in fact it has been demonstrated that there are bipartite entangled mixed states whose correlations admit a classical local model\cite{werner1989PRA}. These NLQCs are unambiguously identified by violations of Bell inequalities\cite{bell,clauser}. These violations have been confirmed by several experiments\cite{aspect1982PRL,mohering2004PRL,groblacher2007Nature} and are crucial for some applications in quantum information\cite{nielsenchuang}, such as to guarantee the safety of device-independent key distribution protocols in quantum cryptography\cite{acin2006PRL,gisin2007natphoton}.

On the other hand, realistic quantum systems present unavoidable interactions with their environment\cite{petru} that, during the evolution, give rise both to decoherence and to entanglement losses, turning any initial pure state into a mixed one. Although these two processes appear to be related, it has been shown that two initially entangled qubits, each locally interacting with a corresponding environment, formed by a zero temperature memoryless (Markovian) bosonic reservoir, may be subjected to ``sudden death''\cite{yu2004PRL}. In this process, the qubits become disentangled at a finite time, in contrast to the single qubit exponential decoherence. This phenomenon has been recently experimentally confirmed\cite{kimble2007PRL,almeida2007Science}. Interest therefore arises in finding physical conditions and methods able to protect entanglement. Recently, methods aimed at extending the entanglement storage time have been proposed. They make use of environments with memory (non-Markovian), where entanglement
revivals occur\cite{bellomo2007PRL,bellomo2008PRA}, of quantum Zeno effect\cite{maniscalco2008PRL}, or of structured environments like photonic crystals, where entanglement can be trapped\cite{bellomo2008arxivtrapping}.

We shall consider a system composed by a couple of qubits, each embedded in a bosonic reservoir, evolving independently to examine the connection between violation of a Bell inequality and single-qubit excited state population. We shall study the special case when the environment is such that sufficiently high population trapping occurs (as in photonic-band gap materials\cite{wolde2003JPB}), to see if a long-time protection of nonlocal entanglement, as identified by Bell inequality violations, is achievable.

\section{Formalism}
We begin by considering a bipartite system made of two identical independent two-level atoms (qubits) each embedded in a zero temperature bosonic environment. We indicate the two qubits with $A$ and $B$, each having ground and excited states $\ket{0}$, $\ket{1}$ separated by a transition frequency $\omega_0$. A crucial point for our study is to obtain the two-qubit reduced density matrix at the time $t$. This can be accomplished by exploiting a procedure based on the knowledge of the single-qubit dynamics\cite{bellomo2007PRL}. The fact that the two parts ``qubit+environment'' evolve independently allows to write the time evolution operator of the composite system as a tensor product of two terms relative to the two parts. This in turn leads to the fact that, given the time-dependent single-qubit density matrix elements as $\rho^{A}_{ii'}(t)=\sum_{ll'}A_{ii'}^{ll'}(t)\rho^{A}_{ll''}(0)$, $\rho^{B}_{jj'}(t)=\sum_{mm'}B_{jj'}^{mm'}(t)\rho^{B}_{mm'}(0)$, the time-dependent two-qubit density matrix elements result of the form\cite{bellomo2007PRL}
\begin{equation}\label{totalevo}
\rho_{ii',jj'}(t)=\sum_{ll',mm'}A_{ii'}^{ll'}(t)B_{jj'}^{mm'}(t)\rho^{}_{ll',mm'}(0),
\end{equation}
where the indexes $i,j,l,m=0,1$. The condition of independent evolution is applicable also to the case of common environment provided that the atoms are at a distance larger than the spatial correlation length of the reservoir and such that the dipole-dipole interaction can be neglected.

For each part, the dynamics of the qubit-environment interaction is described by the Hamiltonian\cite{petru}
\begin{equation}\label{Hamiltonian}
\hat{H}=\hbar \omega_0 \hat{\sigma}_+\hat{\sigma}_-+\sum_k \left[\hbar \omega_k \hat{b}_k^\dag \hat{b}_k+\left(g_k \hat{\sigma}_+\hat{b}_k+g_k^* \hat{\sigma}_- \hat{b}_k^\dag\right)\right],
\end{equation}
where $\sigma_+=\ket{1}\bra{0}$, $\sigma_-=\ket{0}\bra{1}$ are the qubit raising and lowering operators, $b_k^\dag$, $b_k$ are the photon creation and annihilation operators and $g_k$ is the coupling constant of the mode $k$ with frequency $\omega_k$.

When the environment is at zero temperature the single-qubit reduced density matrix $\hat{\rho}^S(t)$ can be written, in the basis $\{\ket{1},\ket{0}\}$, as\cite{petru}
\begin{equation}\label{roS}
\hat{\rho}^S(t)=\left(%
\begin{array}{cc}
\rho^S_{11}(0)|q(t)|^2  & \rho^S_{10}(0)q(t)\\\\
\rho^S_{01}(0)q^*(t)  & \rho^S_{00}(0)+ \rho^S_{11}(0)(1-|q(t)|^2) \\
\end{array}\right),
\end{equation}
where $p(t)=\rho^S_{11}(0)|q(t)|^2$ represents the single-qubit population of excited state at time $t$ and $|q(t)|^2$ is the population parameter. The previous equation shows that the single-qubit dynamics depends only on $q(t)$. The two-qubit density matrix $\hat{\rho}(t)$ is then obtained by means of Eq.~(\ref{totalevo}), its elements depending only on their initial values and on the function $q(t)$. In the following these density matrix elements will be meant in the standard two-qubit basis $\mathcal{B}=\{\ket{1}\equiv\ket{11},\ket{2}\equiv\ket{10}, \ket{3}\equiv\ket{01}, \ket{4}\equiv\ket{00}\}$.

\section{CHSH-Bell inequality}
In order to individuate when the entanglement is nonlocal we have to choose a Bell inequality and look for its violations. Here we shall exploit the well-known Clauser-Horne-Shimony-Holt (CHSH) form of the Bell inequality\cite{clauser}. The CHSH-Bell inequality is the most extensively used among the Bell inequalities. The reason of this relies on the fact that it is experimentally testable\cite{clauser,aspect1982PRL} and that for the case of two spin-like measurements (two measurements with two outcomes per each) any other Bell inequality is equivalent to the CHSH-Bell one\cite{fine1982PRL}.

Let the operator $\mathcal{O}_S=\mathcal{O}_S(\theta_S,\phi_S)$ be a spin-like observable with eigenvalues $\pm1$ associated to the qubit $S=A,B$, defined as $\mathcal{O}_S=\textbf{O}_S\cdot\bm{\sigma}_S$, where $\textbf{O}_S\equiv(\sin\theta_S\cos\phi_S,\sin\theta_S\sin\phi_S,\cos\theta_S)$ is the unit vector indicating a direction in the spin-like space and $\bm{\sigma}_S=(\sigma_1^S,\sigma_2^S,\sigma_3^S)$ the Pauli matrices vector. The expression of $\mathcal{O}_S$ in terms of the basis states is
\begin{eqnarray}
\mathcal{O}_S&=&\cos\theta_S(\ket{1}\bra{1}-\ket{0}\bra{0})\nonumber\\
&+&\sin\theta_S(e^{i\phi_S}\ket{1}\bra{0}+e^{-i\phi_S}\ket{0}\bra{1}).
\end{eqnarray}
The CHSH-Bell inequality associated to the two-qubit state $\hat{\rho}$ for the operator $\mathcal{O}$ is\cite{clauser}
\begin{equation}\label{bellfunction}
B(\hat{\rho})=|\ave{\mathcal{O}_A\mathcal{O}_B}-\ave{\mathcal{O}_A\mathcal{O}'_B}|
+\ave{\mathcal{O}'_A\mathcal{O}_B}+\ave{\mathcal{O}'_A\mathcal{O}'_B}\leq2,
\end{equation}
where $B(\hat{\rho})$ is the Bell function, $\ave{\mathcal{O}_A\mathcal{O}_B}=\mathrm{Tr}\{\hat{\rho}\mathcal{O}_A\mathcal{O}_B\}$ is the correlation function and $\mathcal{O}'_S=\mathcal{O}_S(\theta'_S,\phi_S)$. If, given the state $\hat{\rho}$, it is possible to find a set of angles $\{\phi_A,\phi_B\}$ and $\{\theta_A,\theta'_A,\theta_B,\theta'_B\}$ such that the CHSH-Bell inequality is violated, that is $B(\hat{\rho})>2$, the correlations are nonlocal and cannot be reproduced by any classical local model.

The Bell function at time $t$ is then obtained as a function of $q(t)$ from the evolved two-qubit state $\hat{\rho}(t)$. Our aim is to individuate CHSH-Bell inequality violation regions characterized by $B(\hat{\rho})>2$. In particular, we consider the case when the two-qubit density matrix at time $t$ has an ``X'' structure, that is having non-zero elements only along the main diagonal and anti-diagonal. This has been shown to happen if the initial two-qubit density matrix has itself an ``X'' structure: in fact this form is preserved during the time evolution due to the Hamiltonian $\hat{H}=\hat{H}_A+\hat{H}_B$, where $\hat{H}_S$ $(S=A,B)$ is the single qubit-environment Hamiltonian of Eq.~(\ref{Hamiltonian})\cite{bellomo2007PRL}. The choice of such initial conditions relies on the fact that, on one hand, it simply leads to analytical solutions and, on the other hand, it includes several initial states of physical interest, as Bell-like (pure) or Werner-like (mixed) states\cite{nielsenchuang,popescu1994PRL}.

The maximum of the Bell function $B(\hat{\rho})$ of Eq.~(\ref{bellfunction}) is found using the standard procedure\cite{gisin1991PLA} at the angles $\{\phi_1,\phi_2\}=\{(k+k')\pi-\delta_{14}-\delta_{23},(k-k')\pi-\delta_{14}+\delta_{23}\}$, where $k,k'$ are integer numbers and $\delta_{14}$, $\delta_{23}$ are respectively the phases of $\rho_{14}(0)$, $\rho_{23}(0)$ and in addition $\{\theta_1,\theta'_1,\theta_2(t),\theta'_2(t)\}=
\{0,\frac{\pi}{2},\arctan\frac{\mathcal{Q}(t)}{|\mathcal{P}(t)|},\pi-\theta_2(t)\}$, and in this case results to be\cite{bellomo2008arxivbell}
\begin{equation}\label{maxBell}
B_\mathrm{max}(\hat{\rho}(t))=2\sqrt{\mathcal{P}^2(t)+\mathcal{Q}^2(t)},
\end{equation}
where
{\setlength\arraycolsep{1.5pt}\begin{eqnarray}\label{PandQ}
\mathcal{P}(t)&=&1-2|q(t)|^2[1+\rho_{11}(0)-\rho_{44}(0)-2\rho_{11}(0)|q(t)|^2],\nonumber\\ \mathcal{Q}(t)&=&2|q(t)|^2(|\rho_{14}(0)|+|\rho_{23}(0)|).
\end{eqnarray}}These equations immediately highlight the relation between Bell function and population parameter $|q(t)|^2$. We notice that this expression for $B_\mathrm{max}(\hat{\rho}(t))$ coincides with that given by the formal Horodecki criterion\cite{horodecki1995PLA}.

\section{Study of CHSH-Bell inequality violations}
In this section we study for which values of $|q(t)|^2$ CHSH-Bell inequality violations occur.

\subsection{\label{initial state}Initial states}
We consider as two-qubit initial states extended Werner-like (EWL) states\cite{bellomo2008PRA} that are mixed states reducing to Werner states\cite{werner1989PRA} or to Bell-like states under certain conditions. In this way we shall be able to analyze the effects of both mixedness and degree of entanglement of the initial states on the dynamics of the maximum of Bell function $B_\mathrm{max}(\hat{\rho}(t))$. Moreover, Werner states are encountered during experimental generations of Bell states \cite{hagley1997PRL,kwiat2001Nature}. EWL states are defined as
\begin{eqnarray}\label{EWLstates}
    \hat{\rho}^\Phi(0)&=&r \ket{\Phi}\bra{\Phi}+\frac{1-r}{4}I_4,\nonumber\\
    \hat{\rho}^\Psi(0)&=&r \ket{\Psi}\bra{\Psi}+\frac{1-r}{4}I_4,
\end{eqnarray}
where $r$ indicates the purity of the initial states, $I_4$ is the $4\times4$ identity matrix and
\begin{eqnarray}\label{Bell-likestates}
\ket{\Phi}=\alpha\ket{01}+\beta e^{i\delta}\ket{10},\quad\ket{\Psi}=\alpha\ket{00}+\beta e^{i\delta}\ket{11},
\end{eqnarray}
are the Bell-like states with $\alpha,\beta$ real and $\alpha^2+\beta^2=1$. The EWL states of Eq.~(\ref{EWLstates}) reduce to the Werner-like states for $\alpha=\pm\beta=1/\sqrt{2}$, that is when their pure part becomes a Bell state. For $r=0$ the EWL states become totally mixed states, while for $r=1$ they reduce respectively to the Bell-like (pure) states $\ket{\Phi},\ket{\Psi}$ of Eq.~(\ref{Bell-likestates}). The EWL states $\hat{\rho}^\Phi(0)$, $\hat{\rho}^\Psi(0)$ are more general than the Werner-like states since they can have a non maximally entangled pure part.

The initial two-qubit density matrix elements for the state $\hat{\rho}^\Phi$ are
\begin{eqnarray}\label{wernerstate1}
\rho^\Phi_{11}(0)&=&\frac{1-r}{4}, \quad
\rho^\Phi_{22}(0)=\frac{1-r}{4}+\beta^2 r,\nonumber\\
\rho^\Phi_{33}(0)&=&\frac{1-r}{4}+\alpha^2 r, \quad
\rho^\Phi_{44}(0)=\frac{1-r}{4} , \nonumber \\
\rho^\Phi_{14}(0)&=& 0,\quad \rho^\Phi_{23}(0)= \alpha \beta e^{i\delta}r,
\end{eqnarray}
while for the state $\hat{\rho}^\Psi$
\begin{eqnarray}\label{wernerstate2}
\rho^\Psi_{11}(0)&=&\frac{1-r}{4} +\beta^2 r, \quad
\rho^\Psi_{22}(0)=\frac{1-r}{4},\nonumber\\
\rho^\Psi_{33}(0)&=&\frac{1-r}{4}, \quad
\rho^\Psi_{44}(0)=\frac{1-r}{4}+\alpha^2 r , \nonumber \\
\rho^\Psi_{14}(0)&=& \alpha \beta e^{i\delta}r,\quad \rho^\Psi_{23}(0)= 0.
\end{eqnarray}
These density matrix elements of the EWL states are such that the resulting density matrix has an ``X'' structure.

\subsection{\label{plots}Nonlocal entanglement protection}
\begin{figure}
\begin{center}
\includegraphics[width=8.25 cm, height=5.0 cm]{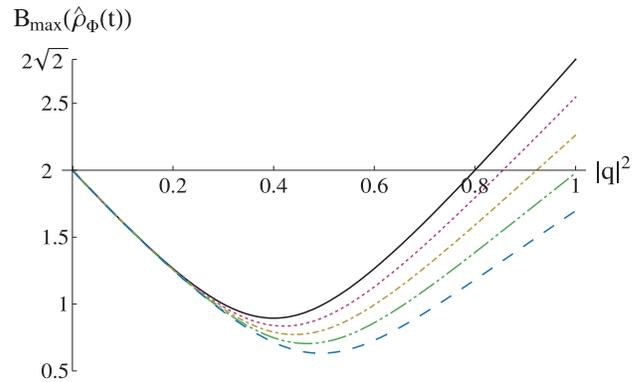}
\caption{\label{fig1:Phi}\footnotesize Maximum of Bell function $B_\mathrm{max}(\hat{\rho}_\Phi(t))$ as a function of the population parameter $|q|^2$ starting from the initial EWL state $\hat{\rho}^\Phi(0)$ with $\alpha=\beta=1/\sqrt{2}$ for different values of purity $r$:  $r=1$ (solid curve), $r=0.9$ (dotted curve), $r=0.8$ (long-short-dashed curve), $r=0.7$ (long-short-short-dashed curve), $r=0.6$ (long-dashed curve).}
\end{center}
\end{figure}

\begin{figure}
\begin{center}
\includegraphics[width=8.25 cm, height=5.0 cm]{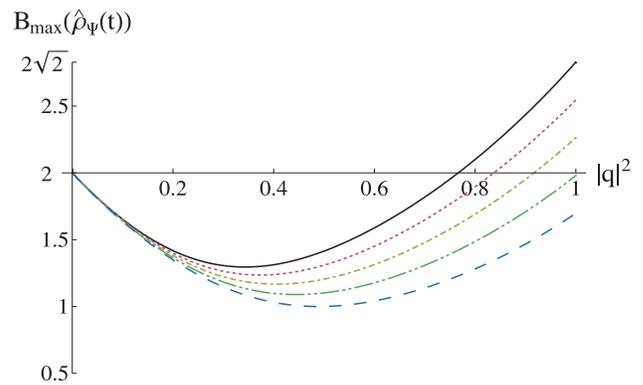}
\caption{\label{fig2:Psi}\footnotesize Maximum of Bell function $B_\mathrm{max}(\hat{\rho}_\Psi(t))$ as a function of the population parameter $|q|^2$ starting from the initial EWL state $\hat{\rho}^\Psi(0)$ with $\alpha=\beta=1/\sqrt{2}$ for different values of purity $r$:  $r=1$ (solid curve), $r=0.9$ (dotted curve), $r=0.8$ (long-short-dashed curve), $r=0.7$ (long-short-short-dashed curve), $r=0.6$ (long-dashed curve).}
\end{center}
\end{figure}

In order to evidence the regions of violations of the CHSH-Bell inequality, we plot the maximums of Bell function $B_\mathrm{max}(\hat{\rho}_\Phi(t))$ and $B_\mathrm{max}(\hat{\rho}_\Psi(t))$, relating to the two EWL states of Eq.~(\ref{EWLstates}), as functions of the population parameter $|q(t)|^2$ for different values of the purity $r$ and for fixed $\alpha$. Fig.~\ref{fig1:Phi} relates to the EWL initial state $\hat{\rho}^\Phi(0)$ while Fig.~\ref{fig2:Psi} to the EWL initial state $\hat{\rho}^\Psi(0)$.
Both figures show that violations of CHSH-Bell inequality can be achieved ($B>2$), for $r$ assigned, when the population parameter $|q(t)|^2$ is larger than a corresponding threshold value $|q_r|^2$. In particular, for initial maximally entangled pure states ($r=1$), the threshold values are $|q_1|^2=0.8$ for the Bell state $\ket{\Phi}=(\ket{01}\pm\ket{10}/\sqrt{2}$ and $|q_1|^2\approx0.77$ for the Bell state $\ket{\Psi}=(\ket{00}\pm\ket{11}/\sqrt{2}$. However, from figures 1 and 2 one sees that for small enough values of purity $r$ ($r<0.71$ for both initial states), CHSH-Bell inequality violations never occur. Therefore, for initial conditions such that the purity of the states is sufficiently high, if physical conditions of the environment are such that high values of single-qubit excited state population can be achieved, nonlocal entanglement between the two qubits is obtained. As a consequence, when these high values of population can be trapped, a long-time protection of nonlocal entanglement follows. These required values of population trapping may be reached, for example, in the case when the environment is constituted by a photonic crystal exhibiting photonic-band gap\cite{wolde2003JPB,bellomo2008arxivtrapping}.

\section{Conclusions}
In this paper, we have considered a system composed by a couple of initially entangled qubits, each embedded in a zero temperature bosonic environment, that evolve independently. In this case, the two-qubit density matrix at time $t$ is simply expressed in terms of the coefficients that determine the evolution of each single independent qubit and on the initial conditions of the qubit pair.

In order to individuate when the two-qubit entanglement presents nonlocal correlations we have exploited the CHSH-Bell inequality and found its relation with the single-qubit excited state population represented by a population parameter. We have graphically shown this relation taking as two-qubit initial conditions Werner-like mixed states. We have found that, when the initial purity of the states is high enough, for some range of the population parameter values, a violation of the CHSH-Bell inequality is achieved. We have suggested that, in the case of structured environments engineered so to present single-qubit population trapping, as for example photonic crystals presenting photonic-band gaps, long-time entanglement protection can be reached.

Nonlocal entanglement means that a Bell inequality is violated and that the two-qubit correlations cannot be reproduced by any classical local model. This condition appears to be relevant both for an applicative point of view, e.g. for quantum information purposes\cite{acin2006PRL,gisin2007natphoton}, and for a fundamental point of view. Our results demonstrate that purely quantum nonlocal correlations can be maintained during the evolution and highlight the importance of using appropriately engineered environments.

R.L.F. (G.C.) acknowledges partial support by MIUR project II04C0E3F3 (II04C1AF4E) \textit{Collaborazioni Interuniversitarie ed Internazionali tipologia C}.

\end{document}